\documentclass[a4paper,showpacs,amsmath,amssymb,12pt]{article}
\usepackage{mathrsfs}
\usepackage{amsmath}
\usepackage{amsfonts}
\usepackage{indentfirst}
\usepackage{amssymb}
\usepackage{latexsym}
\usepackage{graphicx}
\usepackage[english]{babel}
\usepackage{epsfig}
\usepackage{subfigure}
\usepackage{multirow}
\usepackage{pstricks}
\usepackage{float}
\usepackage[colorlinks=true, linkcolor=blue, citecolor=blue, urlcolor=blue]{hyperref}

\title{Indirect production of doubly charmed tetraquark $T_{cc}$ at high energy colliders}
\author{Juan-Juan Niu\footnote{Contact author: niujj@gxnu.edu.cn}, Bin-Bin Shi, Zheng-Kui Tao and Hong-Hao Ma\footnote{Contact author: mahonghao@pku.edu.cn} \\
{\small Guangxi Key Laboratory of Nuclear Physics and Technology,}\\
{\small Department of Physics, Guangxi Normal University, Guilin 541004, China}}
\date{}

\begin{document}
\maketitle
\begin{abstract}
The indirect production mechanisms of doubly charmed tetraquark $T_{cc}$ through three decay channels, Higgs$/Z^{0}\to \langle cc\rangle_{\bar{3}} +\bar{c}+\bar{c} \to T_{cc}^{\bar{q}\bar{q^{\prime}}} +\bar{c}+\bar{c} $ and $W^+\to \langle cc\rangle_{\bar{3}}+\bar{c}+\bar{s} \to T_{cc}^{\bar{q}\bar{q^{\prime}}} +\bar{c}+\bar{s} $, are analyzed within the framework of nonrelativistic QCD. The intermediate $\langle cc\rangle_{\bar{3}}$ diquark cluster in color antitriplet evolves into tetraquark components via the fragmentation process by trapping two light antiquarks ($\bar{q}$ and $\bar{q^{\prime}}$) from the vacuum. After the considered doubly charmed tetraquark components are summed, including $T_{cc}^{\bar{u} \bar{u}}$, $T_{cc}^{\bar{u} \bar{d}}$, $T_{cc}^{\bar{d} \bar{d}}$, $T_{cc}^{\bar{u}\bar{s}}$, and $T_{cc}^{\bar{d}\bar{s}}$, the decay widths, branching ratios, and produced events each year for the production of $T_{cc}$ can be predicted at LHC and CEPC, respectively. The differential distributions and two main sources of theoretical uncertainty are also discussed. The results show that the produced events each year for $T_{cc}$ via $W^{+}$ decays is $1.80\times10^5$, nearly $2$ orders of magnitude larger than that by Higgs decays ($1.11\times10^{3}$) and $Z^{0}$ decays ($4.81\times10^3$) at LHC. However at CEPC, the largest contribution for the production of $T_{cc}$ is through $Z^{0}$ decays, about $1.63\times10^6$. There are only $4.79\times10^{-1}$ and $2.03\times10^{2}$ $T_{cc}$ events produced each year at CEPC through Higgs and $W^+$ decay, respectively.
\end{abstract}

\section{INTRODUCTION}

The quark model, proposed independently by Gell-Mann and Zweig in 1964 \cite{GellMann:1964nj,Zweig:1981pd,Zweig:1964jf}, systematically classifies hadrons into mesons, baryons, and exotics. Among them, exotics have more complex structures, and include tetraquarks, pentaquarks, hybrids, and glueballs. 
Since the observation of X(3872) in 2003, experiments and theories have been devoted to finding and studying exotics and their properties. A review and references of relevant experimental and theoretical status and advances are available~\cite{Brambilla:2019esw,Manohar:1992nd,Maiani:2004vq}. Many new exotics with heavy quarks, such as $T_{cc}$ and $P_{c}$, have been systematically studied in $BABAR$, Belle, and LHCb experiments~\cite{LHCb:2021auc}, and they exhibit behaviors and properties different from those traditional hadrons predicted by the standard model. The doubly charmed tetraquark $T_{cc}^{\bar{q}\bar{q^{\prime}}}$ contains double heavy charm quarks and double light antiquarks $\bar{q}$ and $\bar{q^{\prime}}$ ($\bar{u}$, $\bar{d}$ or $\bar{s}$), specifically $T_{cc}^0$ ($T_{cc}^{\bar{u}\bar{u}}$), $T_{cc}^+$ ($T_{cc}^{\bar{u}\bar{d}}$), $T_{cc}^{++}$ ($T_{cc}^{\bar{d}\bar{d}}$), $T_{cc}^{\bar{u}\bar{s}}$, $T_{cc}^{\bar{d}\bar{s}}$, and $T_{cc}^{\bar{s}\bar{s}}$. For simplicity, unless otherwise stated, we will omit the light antiquark and use $T_{cc}$ to represent the doubly charmed tetraquark. Theoretical methods have been used to deeply understand these tetraquarks, not only the phenomenological extensions of the quark model \cite{Fontoura:2019opw,Mutuk:2024vzv}, but also the chiral perturbation theory~\cite{Leutwyler:1993iq}, QCD sum rules~\cite{Navarra:2012zz} and lattice QCD~\cite{Collins:2024sfi}.

The nonrelativistic QCD (NRQCD)~\cite{Bodwin:1994jh,Petrelli:1997ge} is a multiscale theory, in which the dynamical scales can be, $m_Q v$, $m_Q v^2$, $\Lambda_{\rm QCD}$, etc., where $v$ is the relative velocity between heavy quarks $Q$ in the c.m. frame and $v \ll 1$. This makes it a potential effective theoretical method to analyze the production mechanism of the doubly charmed tetraquark state. In the framework of NRQCD, the double charm quarks inside $T_{cc}$ move in a small relative
velocity $v$ in the rest frame with a larger scale $m_cv$ comparing to the scale of the light degree of freedom ($\Lambda_{\rm QCD}$). For the structure of $T_{cc}$, the perturbatively produced heavy charm quark pair $cc$  through Higgs, $W^+$, or $Z^0$ decay can first form a color-antitriplet diquark cluster $\langle cc\rangle_{\bar{3}}$ by the attractive strong interactions with a mass of approximately 3.6 GeV. Then the color-antitriplet cluster acting as a heavy antiquark supports the color interactions to the other two light antiquarks, just like the single heavy baryon.  For the decomposition of $\rm SU(3)_C$ color group $\mathbf 3\otimes \mathbf 3 = \bar{\mathbf 3}\oplus \mathbf 6$, the color quantum number for the $\langle cc\rangle$ diquark state can be color-antitriplet $\bar{\mathbf 3}$ or color-sextuplet $\mathbf {6}$. However, the diquark $\langle cc\rangle_{\mathbf {6}}$ is not discussed here because the color interaction between this kind of diquark is repulsive~\cite{Maiani:2004vq} and cannot become a kind of cluster within the tetraquark to a large extent. Some theoretical studies on these states were presented in the literatures \cite{Zouzou:1986qh,Vijande:2007rf,Ma:1991wg,Vijande:2009kj,Chen:2011jtl,Qin:2020zlg,Ali:2018xfq,Mutuk:2023oyz,Cheng:2020wxa,Cheng:2022qcm,Montesinos:2023qbx,Braaten:2022elw,Tiwari:2021iqu,Lu:2020rog,Yang:2019itm,Hernandez:2019eox}.  In addition to compact diquark-antidiquark states, there are other alternative dynamical mechanisms to describe the tetraquark state, like meson molecule \cite{Tornqvist:1993ng,Guo:2013sya,Wang:2013daa,Guo:2017jvc} or a hadroquarkonium \cite{Dubynskiy:2008mq,Ferretti:2020ewe}.

The Large Hadron Collider(LHC) and Circular Electron-Positron Collider(CEPC) with high energy and high luminosity provide good experimental platforms for the research of tetraquarks. A previous study on the direct hadronic production of tetraquarks in the subprocess gluon-gluon fusion at LHC has been presented in Ref.~\cite{Chen:2011jtl}. The contributions from indirect production mechanisms are also not negligible \cite{Niu:2018ycb,Niu:2019xuq,Zhang:2022jst,Tian:2023uxe} and they could be an auxiliary force in the search for new hadrons. In this paper, we are interested in the indirect production mechanisms of doubly charmed tetraquark $T_{cc}$ at LHC and CEPC. They can be indirectly produced through three typical high-energy physical decay processes, Higgs$/Z^{0}/W^{+} \to  \langle cc\rangle_{\bar{3}} +\bar{c}+\bar{c}/\bar{s} \to T_{cc}+\bar{c}+\bar{c}/\bar{s}$, where the spin quantum number of the $\langle cc\rangle_{\bar{3}}$ diquark cluster can be $^3 S_1$ and its orbital excited states for the antisymmetry by exchanging identical fermions. 

These three decay processes all can be factorized into the convolution of the perturbative short-distance coefficients and the nonperturbative long-distance matrix elements. In the perturbation area, four free quarks or antiquarks are produced by the decay of the Higgs, $Z^{0}$, and $W^{+}$ bosons, respectively. Then in the nonperturbative region, the resulting two charm quarks form a diquark cluster $\langle cc\rangle_{\bar{3}}$ with certain transition probabilities.  Finally, the $\langle cc\rangle_{\bar{3}}$ diquark cluster evolves into $T_{cc}$ via the fragmentation process by trapping two light antiquarks from the vacuum. Theoretical predictions for the produced events and kinematic behaviors of $T_{cc}$ would be given at LHC and CEPC, respectively. This will provide some theoretical guidance for the experimental follow-up discovery of tetraquark states.

The rest of this paper is arranged as follows. In Sec. \ref{sec2},
the calculation technology for the production of these three subprocesses, Higgs$/Z^{0} \to T_{cc}
+\bar{c}+\bar{c} $ and $W^{+} \to T_{cc} +\bar{c}+\bar{s} $, is presented in NRQCD theory. Then we give the numerical results of the decay widths, produced events, differential distributions, and theoretical uncertainties of $T_{cc}$ at LHC and CEPC in Sec. \ref{sec3}. 
Section \ref{sec4} is devoted to a summary.

\section{CALCULATION TECHNOLOGY}\label{sec2}
~~~~Within the framework of NRQCD, the decay widths of these three subproesses for the production of tetraquark $T_{cc}$ can be factorized into three parts related to three different energy scales ($m_c$, $m_cv$ and $\rm \Lambda_{QCD}$):
\begin{eqnarray}
\Gamma(Higgs/Z^{0}/W^{+}(p_0) &\rightarrow& T_{cc}(p_1)+ \bar {c} (p_2)+\bar {c}/\bar{s}(p_3) )  \nonumber\\
 =\hat{\Gamma}(Higgs/Z^{0}/W^{+} &\rightarrow& \langle cc \rangle_{\bar{3}} + \bar {c} +\bar {c}/\bar{s} ) |\Psi_{cc}(0)|^2\int_{0}^{1}dx D _{\langle cc\rangle_{\bar{3}}  \to T_{cc}}(x),
\label{totwidth}
\end{eqnarray}
where $D_{\langle cc\rangle_{\bar{3}}  \to T_{cc} }(x)$ is
the fragmentation function of the $\langle cc\rangle_{\bar{3}} $ diquark cluster into the color-singlet
tetraquark $T_{cc}$ nonperturbatively. The certain transition probability from the perturbative quark pair $cc$ to the diquark cluster $\langle cc \rangle_{\bar{3}}$ is unknown yet. However, for the $\langle cc \rangle[^3 S_1]_{\bar{\mathbf 3}}$, it can be approximately associated with the Schr\"{o}dinger wave function at the origin $|\Psi_{cc}(0)|^2$ by assuming the potential of the binding color-antitriplet diquark state is hydrogenlike.
And its numerical value of $|\Psi_{cc}(0)|^2$ can be estimated by fitting the experimental data or some nonperturbative methods like QCD sum rules~\cite{Kiselev:1999sc}, lattice QCD~\cite{Bodwin:1996tg}, or the potential model~\cite{Bagan:1994dy}. 

The decay width $\hat{\Gamma}$ in Eq.~(\ref{totwidth}) stands for the perturbative short-distance coefficient describing the production of the $\langle cc \rangle$ diquark state, which can be written as
\begin{eqnarray}
\hat{\Gamma}(Higgs/Z^0/W^+ \rightarrow \langle cc \rangle_{\bar{3}} +\bar {c} +\bar {c}/\bar {s})= \int \frac{1}{2p_{0}^{0}} \overline{\sum} |\mathcal{M}|^2 d\Phi_3,
\label{width}
\end{eqnarray}
in which $\mathcal{M}$ is the hard amplitude, $\overline{\sum}$ means to average over the spin and color of the initial boson and sum over the spins and colors of all the final-state particles, and $d\Phi_3$ is the three-body phase space and can be expressed as
\begin{eqnarray}
d\Phi_3=(2\pi)^4 \delta^4(p_0-\sum_{f=1}^{3} p_f) \prod_{f=1}^{3} \frac{d^3 p_f}{(2\pi)^3 2 p_{f}^{0}}\nonumber.
\end{eqnarray}

After integrating over $d\Phi_3$, the numerical results of the decay widths, the estimated produced events, and the corresponding differential distributions for the indirect production of $T_{cc}$ can be derived at LHC and CEPC, respectively.

\subsection{Amplitude}

\begin{figure}
\centering
  \hspace{-.30in}
 \subfigure[]{
    \includegraphics[scale=0.3]{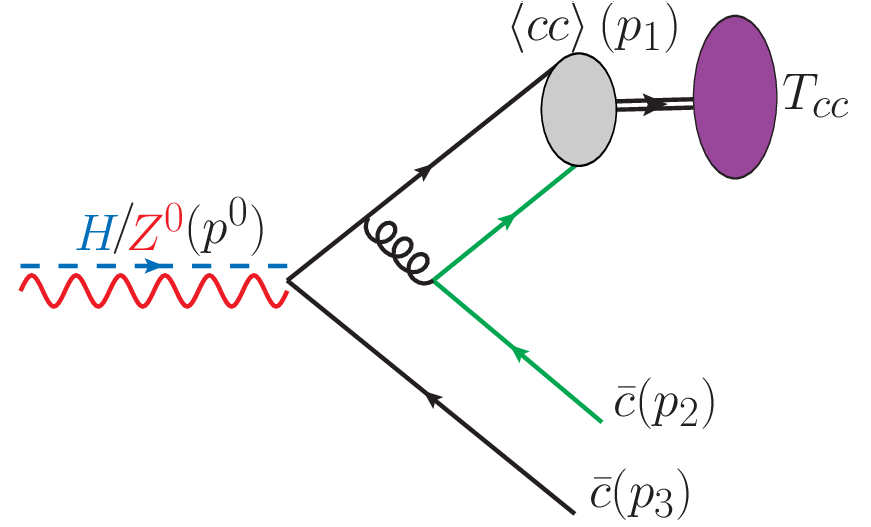}}
  \hspace{-.20in}
  \subfigure[]{
    \includegraphics[scale=0.3]{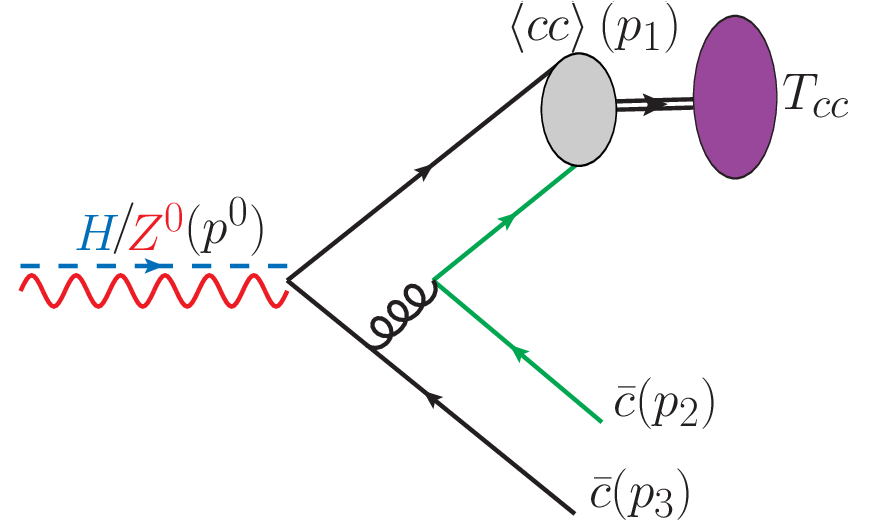}}
  \hspace{-.20in}
  \subfigure[]{
    \includegraphics[scale=0.3]{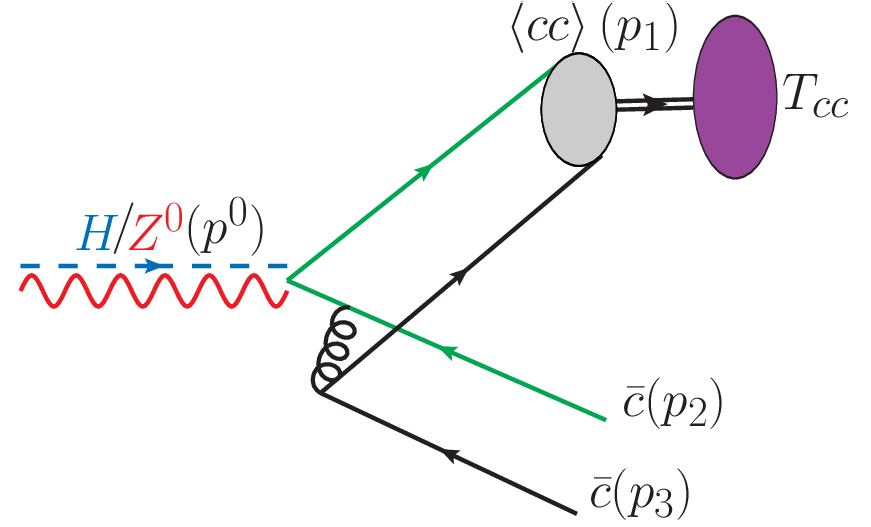}}\\
  \hspace{-.20in}
  \subfigure[]{
    \includegraphics[scale=0.3]{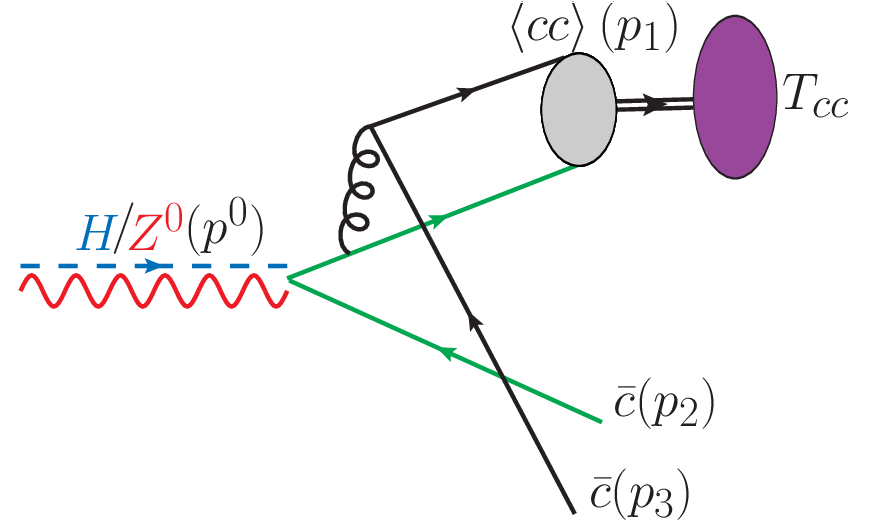}}
  \hspace{-.20in}
   \subfigure[]{
    \includegraphics[scale=0.3]{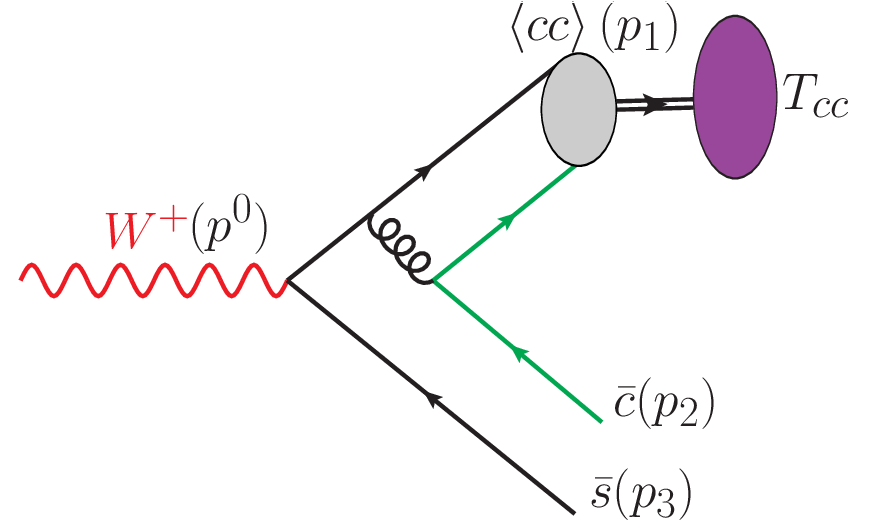}}
  \hspace{-.20in}
  \subfigure[]{
    \includegraphics[scale=0.3]{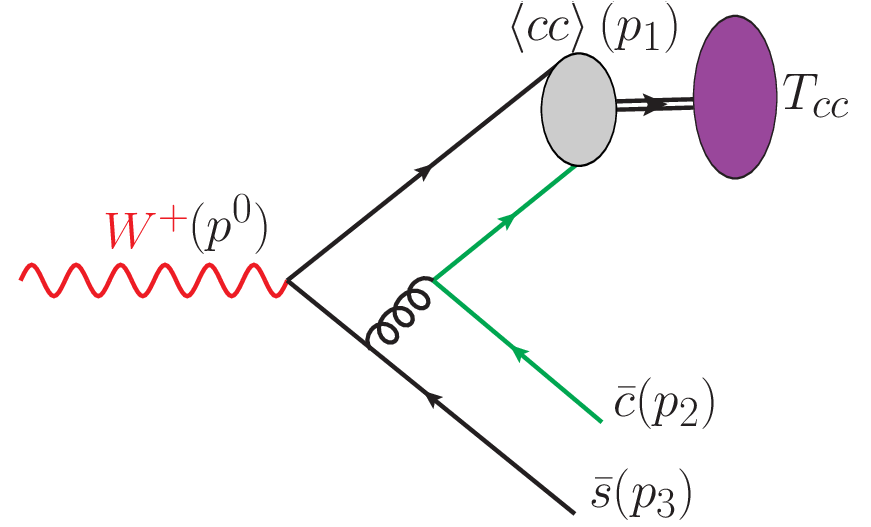}}
\caption{Typical Feynman diagrams for the production of $T_{cc}$ through Higgs and $Z^{0}$ bosons decay [panels (a)-(d)) and $W^{+}$ boson decays [panel (e) and (f)].} \label{fm}
\end{figure}

The typical Feynman diagrams for the production of $T_{cc}$ through Higgs/$Z^{0}/W^{+}$ bosons decay are presented in Fig.~\ref{fm}. At the tree level, the production processes through Higgs and $Z^{0}$ decay involve  four Feynman diagrams [Figs.~\ref{fm}(a)-\ref{fm}(d)] and include two Feynman diagrams via $W^{+}$ boson decays [Figs. \ref{fm}(e) and \ref{fm}(f)].
However, for the production of $\langle cc \rangle$ and $\langle bb\rangle$ diquarks, there are four (two) additional diagrams through Higgs and $Z^0$ $(W^+)$ decay from the exchange of two identical quark lines inside the diquark. It is worth mentioning that the contribution from the decay channel $W^+ \to \langle cc\rangle[n]+ \bar{c} +\bar{b}$ is too small to be considered due to the suppression of the CKM matrix element $|V_{cb}|$ ($|\frac{V_{cb}}{V_{cs}}|^2<0.002$). 

The hard amplitude $\mathcal M$ for the process of $\langle cc\rangle_{\bar{3}}+\bar{c}+\bar{c}/\bar{s}$ can be related to the familiar meson production process $ c\bar{c}+c +\bar{c}/\bar{s}$ with the action of $C$ parity on one of the heavy charm fermion lines (green line in Fig.~\ref{fm}), which has been proven in Refs.~\cite{Zheng:2015ixa,Jiang:2012jt}. Using the charge conjugation matrix $C=-i\gamma^2\gamma^0$, the heavy charm fermion line can be reversed with an additional factor $(-1)^{n+1}$, where $n$ depends on the number of vector vertices on the reversed fermion line, and here $n=1$. Because of the symmetry of identical particles in the $\langle cc\rangle_{\bar{3}}$ diquark, an extra factor $\frac{2^2}{2!}$ has to be multiplied, where the 1/2! factor is for the wave function of identical particles in the diquark and the $2^2$ factor is because there are twice as many diagrams coming from the exchange of the two identical quark lines inside the diquark. Another factor $\frac{1}{2!}$ that needs to be taken into account is the identical particles $\bar{c}$ in the final state through Higgs and $Z^{0}$ boson decay. A detailed expression of the amplitude for the production of  diquark cluster $\langle cc\rangle[^3S_1]_{\bar{\mathbf 3}}$ diquark states can be found in the literature~\cite{Niu:2019xuq,Zhang:2022jst,Tian:2023uxe} through Higgs, $W^{+}$, and $Z^{0}$ bosons decay, correspondingly.

To ensure the gauge invariance, $m_{\langle cc\rangle}=2m_c$ is adopted, $p_{11}$ and $p_{12}$ are the specific momenta of these two constituent quarks inside the diquark:
$p_{11} = \frac{1}{2} p_1 + q$ and $p_{12} = \frac{1}{2} p_1 -q$, 
where $q$ is the relative momentum between these two constituent quarks and it is small enough to be
neglected in the amplitude of the $S$-wave for the nonrelativistic approximation.

\subsection{Fragmentation}

In the heavy quark limit, the fragmentation function for the evolution of the diquark cluster $\langle cc\rangle_{\bar{3}} $ into tetraquark $T_{cc}$ can be approximated by that of a heavy quark $c$ into the single charmed baryon for the same color source. And it can be described nonperturbatively by certain phenomenological models, such as the Peterson model~\cite{Peterson:1982ak} and Bowler model \cite{Bowler:1981sb}. In the following discussion, we will adopt the Peterson model, in which the fragmentation function is expressed as follows:
\begin{eqnarray}\label{frag-sub}
D_{\langle cc\rangle_{\bar{3}}  \to T_{cc}}(x)=\frac{N}{x [ 1-(1/x)- \epsilon_{cc}/(1-x) ]^2}.
\label{fragmentation}
\end{eqnarray}

There are two parameters in Eq.~(\ref{fragmentation}): $\epsilon_{cc}$ and $N$, where $\epsilon_{cc}$ is the parameter determined by experiment. $\epsilon_{b}$ has been determined approximately to be 0.003$\sim0.006$ \cite{SLD:2002poq,ALEPH:2001pfo}. In the Peterson model, there is a scaling behavior for the parameter $\epsilon_{Q}$ that is proportional to $1/m_Q^2$. Based on such a situation, $\epsilon_{cc}$ can be calculated to be $\epsilon_{cc}=(\frac{m_b}{m_{cc}})^2\epsilon_{b}\simeq 0.0082$ with $\epsilon_{b}$=0.004; $N$ is the normalization constant that depends on the total fragmentation probability ($R$) from the $\langle cc\rangle_{\bar{3}} $ diquark cluster to the tetraquark, satisfying the following formula:
\begin{eqnarray}\label{nor-sub}
 \int
dx \,D_{\langle cc\rangle_{\bar{3}}  \to T_{cc}}(x)  =R.
\end{eqnarray}

However, this fragmentation probability $R$ has not been experimentally measured; it can be approximated by analogy with the fragmentation probability of $c\rightarrow \Lambda_{c}^+$ for the production of $T_{cc}^+$, which has been measured at the $pp$ collision and in the $e^+e^-$ collision. Thus, the fragmentation probability to produce the tetraquark $T_{cc}^{+}$ is set to be $R(c\to\Lambda_c^+)=0.168$ at LHC  \cite{ ALICE:2023sgl} and $R(c\to\Lambda_c^+)=0.06$ at CEPC \cite{ Lisovyi:2015uqa}. The 
corresponding normalization constants are calculated to be 0.0268 and 0.0096 at LHC and CEPC, respectively. In Table~ \ref{fp}, we also listed the fragmentation probability $R(c \rightarrow b_c)$ of a charm-quark evolving into a different single charmed baryon $b_c$, such as $\Lambda_{c}^{+}$, $\Xi_c^{0,+}$, and $\Sigma_{c}^{0,+,++}$, at $pp$ collision \cite{ALICE:2023sgl}, $e^{+}e^{-}$ collision \cite{Lisovyi:2015uqa}, and $ep$ collision \cite{Lisovyi:2015uqa}, respectively. Table~ \ref{fp} shows that the fragmentation probability $R(c\to\Lambda_c^+)$ at $e^{+}e^{-}$ and $ep$ collisions are $6.0\%$ and $5.4\%$. As for $R(c\to\Xi_c^{0,+})$, the contribution from $e^{+}e^{-}$ and $ep$ collisions can be negligible compared 
to that at LHC. The fragmentation probability $R(c\to\Sigma_{c}^{0,+,++})$ is to be $7.2\%$, about $6$ times larger than that at the $e^{+}e^{-}$ collision ($1.2\%$) and $ep$ collision ($1.3\%$) estimated for $\Sigma_{c}^{0,+,++}/D^0\approx0.02$ in Refs.~\cite{ALICE:2021rzj,ALICE:2023sgl}. Based on this situation, the decay widths and produced events of $T_{cc}^0$ ($T_{cc}^{\bar{u}\bar{u}}$), $T_{cc}^+$ ($T_{cc}^{\bar{u}\bar{d}}$), $T_{cc}^{++}$ ($T_{cc}^{\bar{d}\bar{d}}$), $T_{cc}^{\bar{u}\bar{s}}$, and $T_{cc}^{\bar{d}\bar{s}}$, can be all estimated at LHC, CEPC, and LHeC. After these doubly charmed tetraquark components are summed, we can predict the decay width and produced events of $T_{cc}$.

Indeed, the fragmentation function is scale dependent, and the initial energy scale was adopted to be $2m_c$ to ensure that gluons are hard enough to split into two charm quarks. The energy scale of the fragmentation function can be evolved into another one by DGLAP equation \cite{DGLAP}. However, at leading order, the fragmentation probability $R$ in Eq.~(4) does not evolve with the factorization scale \cite{Niu:2018ycb,Braaten:1993jn}, which is what we focus on in this paper. Therefore, in the numerical calculations we can take the fragmentation functions with the initial energy scale to be the mass of $T_{cc}$ without evolution.

\begin{table}[htb]
\begin{center}
\caption{Fragmentation probability $R$ of a charm quark evolving into a single charmed baryon at $pp$, $e^{+}e^{-}$, and $ep$ collisions, respectively.} \vspace{0.5cm}
\begin{tabular}{|c|c|c|c|}
\hline
 $R(c \rightarrow b_c)$& $pp$ Collision  & $e^{+}e^{-}$ Collision  & $ep$ Collision\\
\cline{2-3}
\hline
$\Lambda_{c}^{+}$ &  16.8 \% &  6.0 \%  & 5.4 \%\\
$\Xi_c^{0}$ &  9.9 \% & - & -\\
$\Xi_c^{+}$ &  9.6 \% & - & -\\
$\Sigma_{c}^{0,+,++}$ &  7.2 \% & 1.2 \% & 1.3 \%\\
\hline
\end{tabular}
\label{fp}
\end{center}
\end{table}

\section{NUMERICAL RESULTS}\label{sec3}
The input parameters in the numerical calculation are listed below:

\begin{eqnarray}
&&m_c=1.8~\rm{GeV},~~~~~~~~~~~~~\it{m_b}=\rm 5.1~{GeV},~~~~~~~\it{m_s}=\rm 0~{GeV},\nonumber\\
&&\it{m_W}=\rm 80.385~{GeV},~~~~~~~\it{m_Z}=\rm 91.1876~\rm{GeV},~\it{m_H}=\rm 125.18~{GeV}, \nonumber\\
&&\Gamma_{\it W}=~{2.085~\rm GeV},~~~~~~~~\Gamma_{\it Z}=2.4952~{\rm GeV},~~~\Gamma_{\it H}=4.2~\rm{MeV},\nonumber\\
&&G_{F}=1.1663787 \times 10^{-5},~\rm{cos} \theta_W=\it m_W / m_Z,~~~|\rm \Psi_{cc}(0)|^2=0.039~{\rm GeV}^3,
\label{parameters}
\end{eqnarray}
where the quark masses and wave function at the origin are the same as Ref.~\cite{Baranov:1995rc} and the others can be obtained from the PDG~\cite{ParticleDataGroup:2020ssz}. The typical energy scale is selected to be $2m_c$ to obtain the strong running coupling $\alpha_s(2m_c)$=0.242. FeynArts 3.9 \cite{Hahn:2000kx}, FeyCalc 9.3 \cite{Shtabovenko:2020gxv}, and the modified FormCalc programs \cite{Hahn:1998yk} are used to make the algebraic and numerical calculations.

The total events of Higgs, $Z^{0}$, and $W^{+}$ ($N_{tot}$) produced at LHC, CEPC, and LHeC are presented in Table \ref{E}. Running at $\sqrt{s}$ = 14 TeV with the luminosity $\mathcal{L} \propto10^{34}~\rm cm^{-2} s^{-1}$, there could be $1.65\times 10^8$ Higgs and $3.07\times10^{10}$ $W^+$, and $1.00\times 10^9$ $Z^0$ bosons produced at LHC each year \cite{Higgsevents,Gaunt:2010pi,Qiao:2011yk,Liao:2015vqa}. As for CEPC \cite{CEPCStudyGroup:2023quu} running at 240 GeV with integrated luminosity 2.2 $\rm ab^{-1}$, a total $4.30\times 10^5$ of Higgs events could be produced each year.
There would be $2.10\times 10^8$ $W^+$ events produced each year at CEPC when the collision energy is 160 GeV and integrated luminosity is 6.9 $\rm ab^{-1}$. As for the total number of events for $Z^0$, there would be $2.05\times 10^{12}$ events produced each year at CEPC when it would be running as a Super-Z factor at collision energy 91 GeV and integrated luminosity 50 $\rm ab^{-1}$. At LHeC with luminosity $\mathcal{O}(100) fb^{-1}$, there would be about $\mathcal{O}(10^4)$ events of Higgs, $\mathcal{O}(10^6)$ events of $W^+$, and $\mathcal{O}(10^5)$ events of $Z^0$ produced in one operation year \cite{LHeCStudyGroup:2012zhm,Baur:1992sj}. However, the number of events at LHeC compared to that at LHC and CEPC is too small to be discussed later in the numerical discussion.

\begin{table}[htb]
\begin{center}
\caption{Total events of Higgs, $Z^0$, and $W^{+}$ produced at LHC, CEPC, and LHeC, respectively.} \vspace{0.5cm}
\begin{tabular}{|c|c|c|c|}
\hline
$N_{tot}$  & LHC  & CEPC \cite{CEPCStudyGroup:2023quu} &LHeC \cite{LHeCStudyGroup:2012zhm,Baur:1992sj}\\
\cline{2-3}
\hline
Higgs & $1.65\times 10^8$\cite{Higgsevents}  & $4.30\times 10^5$ & $\mathcal{O}(10^4)$ \\
$W^+$ & $3.07\times 10^{10}$\cite{Qiao:2011yk}  & $2.10\times 10^8$  & $\mathcal{O}(10^6)$ \\
$Z^0$ & $1.00\times 10^9$\cite{Liao:2015vqa} & $2.05\times 10^{12}$  & $\mathcal{O}(10^5)$ \\
\hline
\end{tabular}
\label{E}
\end{center}
\end{table}

\subsection{Decay width}

The predicted decay widths, branching ratios, and produced events of $T_{cc}^+$ at LHC and CEPC are all estimated through three decay channels, Higgs$/Z^{0}/W^{+}\rightarrow T_{cc}^+ + \bar {c} +\bar {c}/\bar{s}$, which are presented in Table \ref{TT}. The produced events of $T_{cc}^+$ are roughly estimated by $N_{tot}\times$Br each year at colliders, in which Br stands for the branching ratios $\frac{\Gamma_{Higgs/W^{+}/Z^{0} \to T_{cc}^++X}}{\Gamma_{Higgs/W^{+}/Z^{0}}}$.

\begin{table}[htb]
\begin{center}
\caption{The predicted decay widths ($\Gamma$), Br, and produced events of $T_{cc}^{+}$ at LHC and CEPC each year, respectively, resulting from Higgs$/Z^{0}/W^{+}\rightarrow T_{cc}^+ + \bar {c} +\bar {c}/\bar{s}$ with the fragmentation probability $R$ is 0.168 at LHC and 0.06 at CEPC. } \vspace{0.5cm}
\begin{tabular}{|c|c|c|c|}
\hline

LHC &  $\Gamma ($GeV)  & Br & Events\\
\hline
$H\to T_{cc}^{+}+X$ & $1.10\times 10^{-8}$ & $2.61\times 10^{-6}$ & $4.30\times 10^2$\\
$W^+\to T_{cc}^{+}+X$ & $4.72\times 10^{-6}$ & $2.27\times 10^{-6}$ & $6.96\times 10^5$\\
$Z^0\to T_{cc}^{+}+X$ & $3.87\times 10^{-6}$ & $1.86\times 10^{-6}$ & $1.86\times 10^3$\\
\hline\hline
CEPC & $\Gamma ($GeV)  & Br  & Events\\
\hline
$H\to T_{cc}^{+}+X$ & $3.91\times 10^{-9}$ & $9.32\times 10^{-7}$ & $4.01\times 10^{-1}$\\
$W^+\to T_{cc}^{+}+X$ & $1.69\times 10^{-6}$ & $8.09\times 10^{-7}$ & $1.69\times 10^2$\\
$Z^0\to T_{cc}^{+}+X$ & $1.38\times 10^{-6}$ & $6.64\times 10^{-7}$ & $1.36\times 10^6$\\
\hline
\end{tabular}
\label{TT}
\end{center}
\end{table}

From Table \ref{TT}, we can see that the decay widths at LHC are relatively larger than that at CEPC, mainly because of the high fragmentation probability $R$ at LHC. The branching ratio through three decay channels is about $\mathcal{O}(10^{-6})$ both at LHC and CEPC. The produced $T_{cc}^+$ events at LHC are much larger than the one at CEPC through Higgs and $W^+$ decay and, in particular, the events produced via Higgs decay at CEPC each year are very small or can be negligible. 
What is surprising is that they would produce quite a large number of $T_{cc}^+$ events through $Z^0$ decays at CEPC, up to $1.36\times10^6$ each year. This is because CEPC produces an abundance of $Z^0$ bosons every year when it runs at the energy of $m_Z$ as a Super-Z factory. 

With the help of the fragmentation probability $c\rightarrow \Lambda_{c}^+$ listed in Table \ref{fp}, we get the fragmentation probability $R$ of $\langle cc \rangle_{\bar{3}} \rightarrow  T_{cc}^+$, as well as the corresponding decay widths, branch ratios, and produced events through three decay channels, and the results are listed in Table \ref{TT}. Similarly, by adding the fragmentation probabilities $R(c \rightarrow b_c)$ of all the different single charmed baryon $b_c$ components listed in Table \ref{fp}, such as $\Lambda_{c}^{+}$, $\Xi_c^{0,+}$, and $\Sigma_{c}^{0,+,++}$, we obtain the total fragmentation probability $R$ of $\langle cc \rangle_{\bar{3}} \rightarrow  T_{cc}$, which can be $R=0.435$ at LHC and $R=0.072$ at CEPC. The theoretical predictions for the production of $T_{cc}$ are shown in Table \ref{TE}.

\begin{table}[htb]
\begin{center}
\caption{The predicted decay widths ($\Gamma$), Br, and produced events of $T_{cc}$, including $T_{cc}^{0}$, $T_{cc}^{+}$, $T_{cc}^{++}$, $T_{cc}^{\bar{u}\bar{s}}$, and $T_{cc}^{\bar{d}\bar{s}}$, at LHC and CEPC each year through three decay channels, Higgs$/Z^{0}/W^{+}\rightarrow T_{cc} + \bar {c} +\bar {c}/\bar{s}$, respectively.} \vspace{0.5cm}
\begin{tabular}{|c|c|c|c|}
\hline
LHC & $\Gamma ($GeV)  & Br  &  Events\\
\hline
$H\to T_{cc}+X$ & $2.84\times 10^{-8}$ & $6.75\times 10^{-6}$ & $1.11\times 10^3$\\
$W^+\to T_{cc}+X$ & $1.22\times 10^{-5}$ & $5.87\times 10^{-6}$ & $1.80\times 10^5$\\
$Z^0\to T_{cc}+X$ & $1.00\times 10^{-5}$ & $4.81\times 10^{-6}$ & $4.81\times 10^3$\\
\hline
\hline
CEPC &  $\Gamma ($GeV)  & Br  &  Events\\
\hline
$H\to T_{cc}+X$ & $4.68\times 10^{-9}$ & $1.11\times 10^{-6}$ & $4.79\times 10^{-1}$\\
$W^+\to T_{cc}+X$ & $2.02\times 10^{-6}$ & $9.68\times 10^{-7}$ & $2.03\times 10^2$\\
$Z^0\to T_{cc}+X$ & $1.66\times 10^{-6}$ & $7.94\times 10^{-7}$ & $1.63\times 10^6$\\
\hline
\end{tabular}
\label{TE}
\end{center}
\end{table}

Table \ref{TE} shows the predicted decay widths, branching ratios, and produced events each year of $T_{cc}$, including $T_{cc}^{0}$, $T_{cc}^{+}$, $T_{cc}^{++}$, $T_{cc}^{\bar{u}\bar{s}}$, and $T_{cc}^{\bar{d}\bar{s}}$, at LHC and CEPC through these three decay channels, Higgs$/Z^{0}/W^{+}\rightarrow T_{cc} + \bar {c} +\bar {c}/\bar{s}$, respectively. The produced events for $T_{cc}$ via $W^{+}$ decays is nearly $2$ orders of magnitude larger than that by Higgs and $Z^{0}$ decay at LHC. However at CEPC, the largest contribution for the production of $T_{cc}$ is through $Z^{0}$ decays and about $1.63\times10^6$ events can be produced each year.

\subsection{Differential distributions}

\begin{figure}
\centering
 \subfigure[]{
    \includegraphics[scale=0.3]{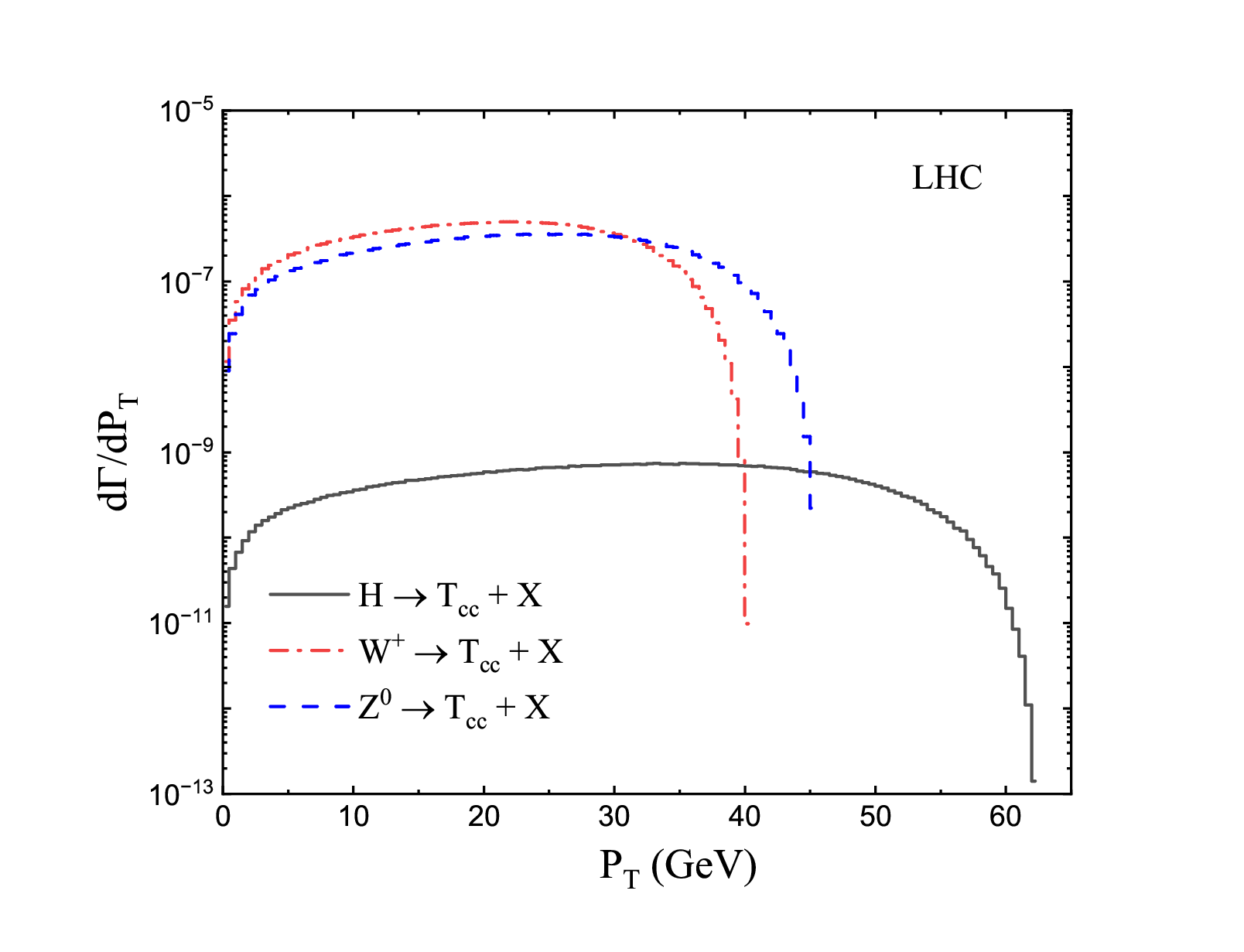}}
  \hspace{-0.70in}
  \subfigure[]{
    \includegraphics[scale=0.3]{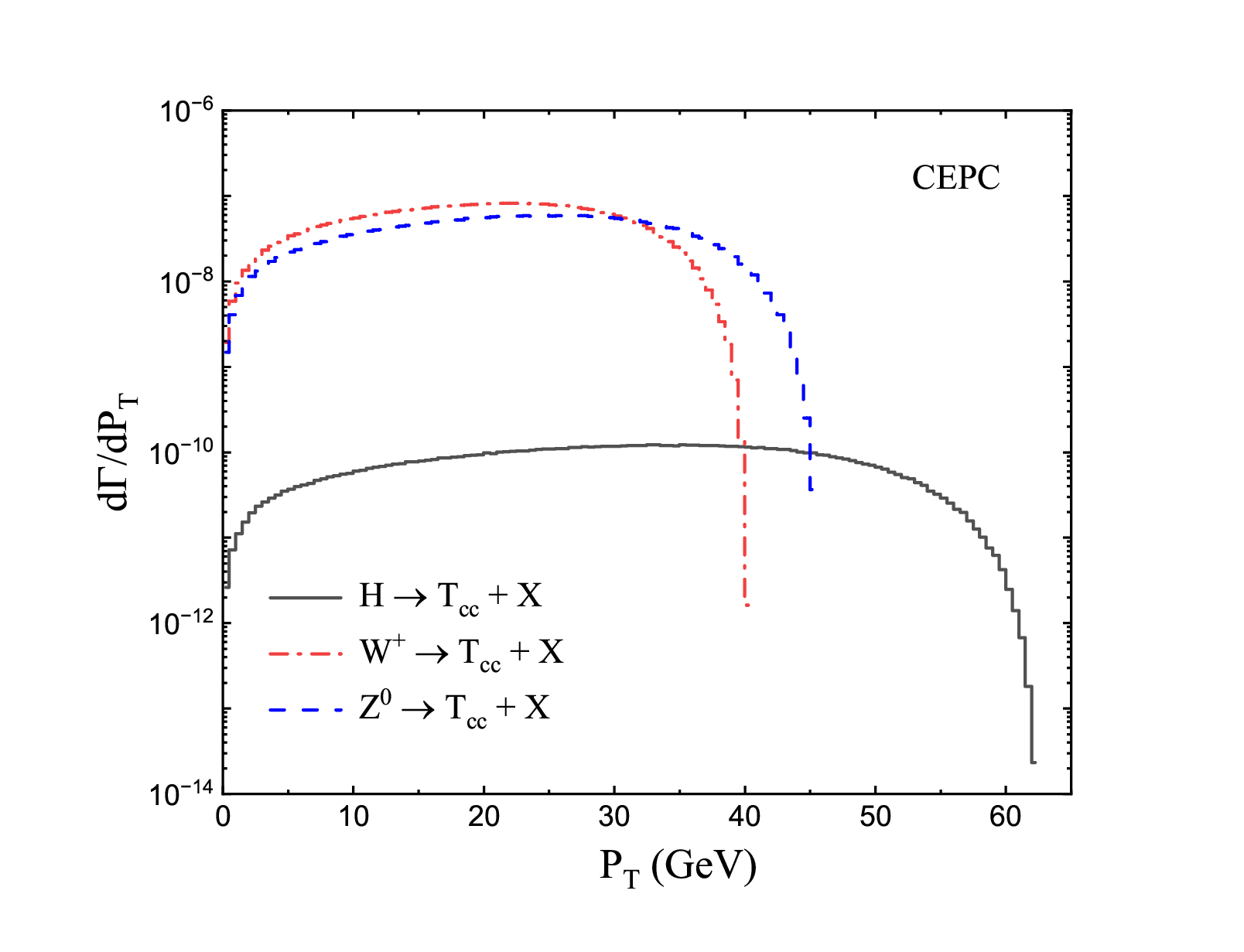}}\\
  \subfigure[]{
    \includegraphics[scale=0.3]{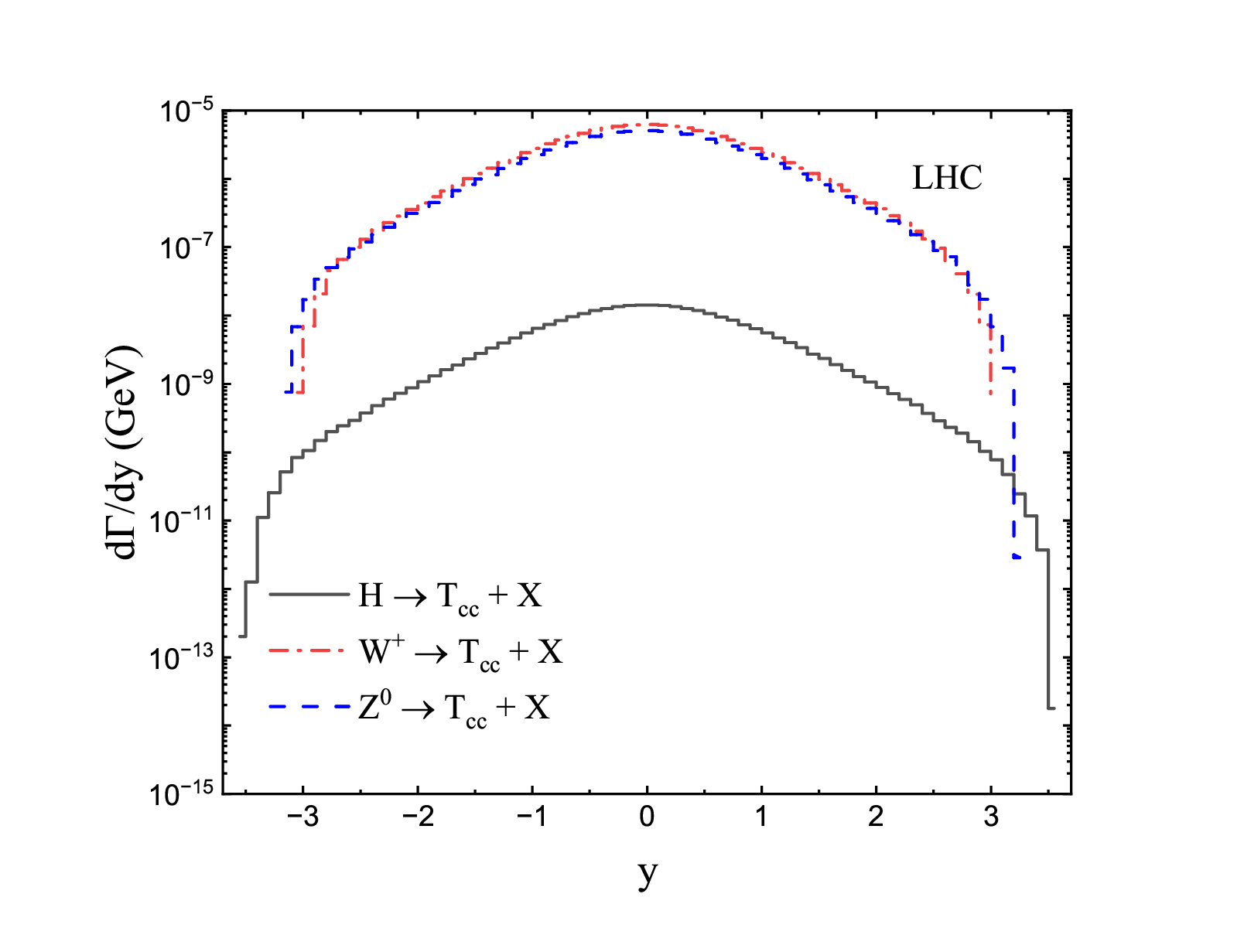}}
  \hspace{-0.70in}
  \subfigure[]{
    \includegraphics[scale=0.3]{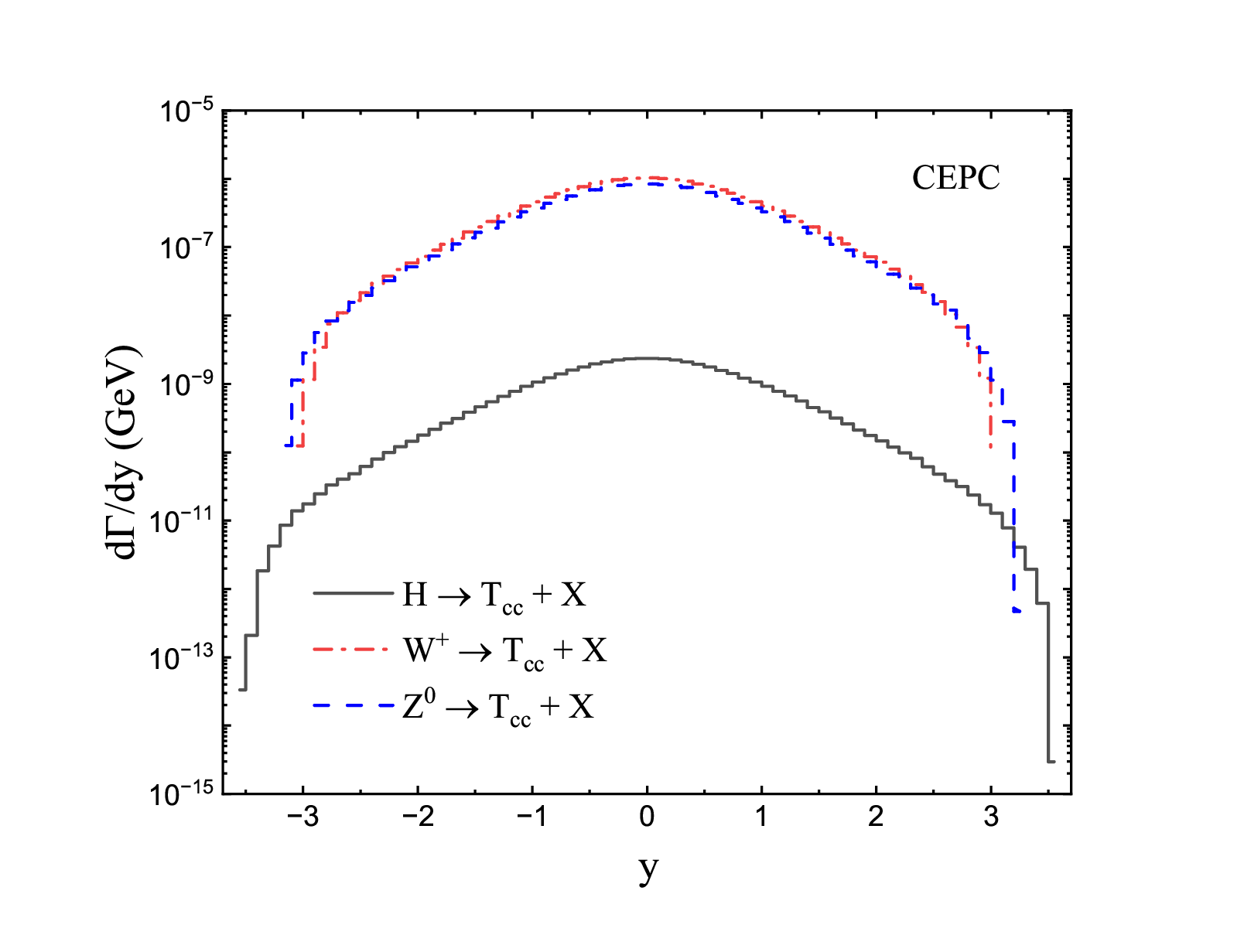}}
\caption{The predicted $P_T-$ and $y-$ distributions of the tetraquark $T_{cc}$ at LHC and CEPC, respectively.} \label{pt_y}
\end{figure}

The transverse momentum $P_T$ distribution and rapidity  $y$  distribution for the production of $T_{cc}$ at LHC and CEPC are shown in Fig.~\ref{pt_y}. Comparing with the $P_T$ and $y$ distributions through $W^+$ and $Z^0$ decay, the differential decay widths via Higgs decays are about $3$ orders of magnitude smaller and the $P_T$ distribution is wider, in the range of $0\sim62.5$~GeV. The behavior of $P_T$ distributions and rapidity distributions through these three decay channels is similar. With the $P_T$ increase, the $P_T$ distributions flatten first and then decrease sharply. For the $y$ distributions, the range is ($-3.6\sim3.6$), and the contributions in the small rapidity distribution interval is greater. These signals will provide some help for the subsequent experiment to find the tetraquark states.

\subsection{Theoretical uncertainty}

There are two main sources of theoretical uncertainty in the numerical calculation, namely, the heavy quark mass $m_c$ and the wave function at the origin $|\Psi_{cc}(0)|^2$ of the $\langle cc\rangle$ diquark.

First, the theoretical uncertainty comes from the mass of heavy quark $m_c$ for building the mass of the diquark. To estimate such uncertainty, we adopt $m_c=1.8\pm0.3~\rm{GeV}$, which is shown in Table~\ref{TU}. 
It is worth mentioning that except for $m_c$, the other parameters listed in Eq.~(\ref{parameters}) need to remain unchanged. Table~\ref{TU} shows that the decay widths for the production of $T_{cc}$ at LHC and CEPC decrease with the increment of $m_c$, which is mainly due to the suppression of phase space. The uncertainty induced by $m_c$ for the production of $T_{cc}$ through $W^+$ and $Z^0$ decay is much larger than that through Higgs decays, mainly due to the interaction of Higgs and the quark pair proportional to $m_c$, and the coupling strength increases with the increase of $m_c$. 
Through Higgs decays at LHC (CEPC), the uncertainty of the decay widths is $\Gamma^{+21.48\%}_{-15.49\%}$ ($^{+21.58\%}_{-15.38\%}$). As for the uncertainty via $W^+$ decays, the range is $\Gamma^{+75.41\%}_{-37.70\%}$ ($^{+74.75\%}_{-38.12\%}$) for LHC (CEPC). And the uncertainty of the decay width is $\Gamma^{+77.00\%}_{-38.20\%}$ ($^{+75.90\%}_{-38.55\%}$) for LHC (CEPC) through $Z^0$ decays.

\begin{table}[htb]
\begin{center}
\caption{Theoretical uncertainty of the decay width $\Gamma$~(GeV) comes from $m_c$ by varying $m_c=1.8\pm0.3~\rm{GeV}$ at LHC and CEPC through three decay channels, Higgs$/Z^{0}/W^{+}\rightarrow T_{cc} + \bar {c} +\bar {c}/\bar{s}$, respectively.} \vspace{0.5cm}
\begin{tabular}{|c|c|c|c|}
\hline
 LHC &1.5  & 1.8  &   2.1\\
\hline
$H\to T_{cc}+X$ & $3.45\times 10^{-8}$ & $2.84\times 10^{-8}$  & $2.40\times 10^{-8}$\\
$W^+\to T_{cc}+X$ & $2.14\times 10^{-5}$ & $1.22\times 10^{-5}$ & $7.60\times 10^{-6}$\\
$Z^0\to T_{cc}+X$ & $1.77\times 10^{-5}$ & $1.00\times 10^{-5}$ & $6.18\times 10^{-6}$\\
\hline
\hline
 CEPC &1.5  & 1.8  &   2.1\\
\hline
$H\to T_{cc}+X$ &  $5.69\times 10^{-9}$ & $4.68\times 10^{-9}$ & $3.96\times 10^{-9}$\\
$W^+\to T_{cc}+X$ & $3.53\times 10^{-6}$ & $2.02\times 10^{-6}$ & $1.25\times 10^{-6}$\\
$Z^0\to T_{cc}+X$ & $2.92\times 10^{-6}$ & $1.66\times 10^{-6}$ & $1.02\times 10^{-6}$\\
\hline
\end{tabular}
\label{TU}
\end{center}
\end{table}

The second theoretical uncertainty arising from the choice of the wave function at the origin is analyzed, whose value is slightly different obtained in different potential models, such as the Power-law~\cite{Bagan:1994dy}, $\rm K^2O$ potential~\cite{Kiselev:2002iy}, and $\rm Buchm\ddot{u}ller$-$\rm Tye~(B.T.)$~\cite{Kiselev:2001fw}. Here for an intuitively comparison we listed them in Table~\ref{WF}, where we can see that the wave function at the origin obtained by the Power-law potential is larger than those calculated under
$\rm K^2O$ and B.T. potential models. To analyze the uncertainty caused by the wave functions at the origin of the $\langle cc\rangle$ diquark, we take the average of these three wave functions at the
origin as the central value, and the uncertainty of the wave functions at the origin is 
$0.0277^{+0.0113}_{-0.0059}$ GeV$^{3}$. The uncertainty of the decay widths for the production of $T_{cc}$ caused by the wave functions at the origin are $2.01^{+0.82}_{-0.43}\times 10^{-8}$ ( $3.33^{+1.35}_{-0.71}\times 10^{-9}$) GeV, $8.69^{+3.54}_{-1.86}\times 10^{-6}$ ($1.43^{+0.58}_{-0.31}\times 10^{-6}$) GeV, and  $7.12^{+2.91}_{-1.53}\times 10^{-6}$ ( $1.18^{+0.48}_{-0.25}\times 10^{-6}$) GeV via Higgs, $W^+$, and $Z^0$ decay at LHC (CEPC), respectively. Fortunately, at the present considered pQCD level, the wave function at the origin can be regarded as a global factor to update the numerical results if more accurate values of the wave function at the origin are available.

\begin{table}[htb]
\begin{center}
\caption{Theoretical uncertainty of the decay width $\Gamma$~(GeV) comes from the wave functions at the origin at LHC and CEPC through three decay channels, Higgs$/Z^{0}/W^{+}\rightarrow T_{cc} + \bar {c} +\bar {c}/\bar{s}$, respectively.} \vspace{0.5cm}
\begin{tabular}{|c|c|c|c|}
\hline
\multirow{2}{*}{LHC}  & Power-law   & $\rm K^2O$ & B.T. \\ \cline{2-4}
 &$|\rm \Psi_{cc}(0)|^2=$0.0390 & $|\rm \Psi_{cc}(0)|^2=$0.0218 & $|\rm \Psi_{cc}(0)|^2=$0.0224\\
\hline
$H\to T_{cc}+X$ & $2.84\times 10^{-8}$ & $1.58\times 10^{-8}$  & $1.63\times 10^{-8}$\\
$W^+\to T_{cc}+X$ & $1.22\times 10^{-5}$ & $6.83\times 10^{-6}$ & $7.01\times 10^{-6}$\\
$Z^0\to T_{cc}+X$ & $1.00\times 10^{-5}$ & $5.60\times 10^{-6}$ & $5.75\times 10^{-6}$\\
\hline
\hline
 \multirow{2}{*}{CEPC} & Power-law   & $\rm K^2O$ & B.T. \\ \cline{2-4}
&$|\rm \Psi_{cc}(0)|^2=$0.0390 & $|\rm \Psi_{cc}(0)|^2=$0.0218 & $|\rm \Psi_{cc}(0)|^2=$0.0224\\
\hline
$H\to T_{cc}+X$ &  $4.68\times 10^{-9}$ & $2.61\times 10^{-9}$ & $2.68\times 10^{-9}$\\
$W^+\to T_{cc}+X$ & $2.02\times 10^{-6}$ & $1.13\times 10^{-6}$ & $1.16\times 10^{-6}$\\
$Z^0\to T_{cc}+X$ & $1.66\times 10^{-6}$ & $9.24\times 10^{-7}$ & $9.48\times 10^{-7}$\\
\hline
\end{tabular}
\label{WF}
\end{center}
\end{table} 


\section{SUMMARY}\label{sec4}

The indirect production mechanisms of the doubly charmed tetraquark are analyzed through three typical decay channels, Higgs$/Z^{0} \to T_{cc}
+\bar{c}+\bar{c} $ and $W^{+} \to T_{cc}
+\bar{c}+\bar{s} $ within the NRQCD framework, where $T_{cc}$ contains the doubly charmed tetraquark components,  $T_{cc}^{0}$, $T_{cc}^{+}$, $T_{cc}^{++}$, $T_{cc}^{\bar{u}\bar{s}}$, and $T_{cc}^{\bar{d}\bar{s}}$. When these states are summed, the decay widths, branching ratios, and produced events for the production of $T_{cc}$ can be predicted at LHC and CEPC each year, respectively. The results show that the produced events for $T_{cc}$ via $W^{+}$ decays is $1.80\times10^5$, nearly $2$ orders of magnitude larger than that by Higgs decays ($1.11\times10^{3}$) and $Z^{0}$ decays ($4.81\times10^3$) at LHC per year. However at CEPC, the largest contribution for the production of $T_{cc}$ is through $Z^{0}$ decays, about $1.63\times10^6$ events produced in one operation year. There are $2.03\times10^{2}$ $T_{cc}$ events produced at CEPC each year through $W^+$ decay. And the events obtained by Higgs decay are too small to be ignored at CEPC each year.

Then $P_T$ and $y$ differential distributions are presented, respectively, at LHC and CEPC for the production of $T_{cc}$. The detailed analysis of differential distribution is helpful to provide some guidance for the subsequent search of the experiment. Comparing with the $P_T$ and rapidity distributions through $W^+$ and $Z^0$ decay, the differential decay widths via Higgs decays are about $3$ orders of magnitude smaller and the $P_T$ distribution is wider, in the range of $0\sim62.5$~GeV. The behavior of $P_T$ distributions and rapidity distributions through these three decay channels are similar. With the $P_T$ increase, the distributions flatten first and then decrease sharply. However, for the rapidity distributions, the contributions in the small rapidity distribution interval are greater.  

Finally, two main sources of theoretical uncertainty, namely, $m_c$ and the wave function at the origin $|\Psi_{cc}(0)|^2$ of the $\langle cc\rangle$ diquark, are discussed respectively at LHC and CEPC.
The decay width for the production of $T_{cc}$ at LHC and CEPC decreases with the increment of $m_c$. The uncertainty induced by $m_c$ for the production of $T_{cc}$ through $W^+$ and $Z^0$ decay is much larger than that through Higgs decays, mainly due to the interaction of Higgs and quark pair proportional to $m_c$. To analyze the uncertainty caused by the wave function at the origin of the $\langle cc\rangle$ diquark, we take the average of three wave functions at the origin obtained by Power-law, $\rm K^2O$, and $\rm Buchm\ddot{u}ller$-$\rm Tye~(B.T.)$ potentials as the central value, and the results are that the uncertainty of the decay widths caused by the wave functions at the origin are $2.01^{+0.82}_{-0.43}\times 10^{-8}$ ( $3.33^{+1.35}_{-0.71}\times 10^{-9}$) GeV, $8.69^{+3.54}_{-1.86}\times 10^{-6}$ ($1.43^{+0.58}_{-0.31}\times 10^{-6}$) GeV,  and $7.12^{+2.91}_{-1.53}\times 10^{-6}$ ( $1.18^{+0.48}_{-0.25}\times 10^{-6}$) GeV for $T_{cc}$ production via Higgs, $W^+$, and $Z^0$ decay at LHC (CEPC), respectively. 

To be mentioned that the subsequent decay of $T_{cc}$ is mainly through the weak decay and semilight decay, $c\to W^+ + s \to l^+ +\nu_l +s$, where the lepton $l^+$ can be $e^{+}$ or $\mu^{+}$, and $\nu_l $ is a corresponding invisible neutrino. In the $T_{cc}$, double $c$ quarks decay in this way
independently, with the two antiquarks as spectators. Considering the signatures and the efficiencies of the the
detectors, we see that high luminosity LHC (HL-LHC) and CEPC provides the best platforms to discover doubly charmed tetraquarks. \\

{\bf ACKNOWLEDGMENTS}\\ This work was partially supported by the Natural Science Foundation of Guangxi (Grant No. 2024GXNSFBA010368) and the Guangxi Technology Base and Talent Subject (Grant No. Guike AD23026182). This work was also supported by the Central Government Guidance Funds for Local Scientific and Technological Development, China (Grant No. Guike ZY22096024) and the National Natural Science Foundation of China (Grant No. 12005045).

\end{document}